\documentclass[onecolumn,12pt]{IEEEtran}
\usepackage[ruled,vlined]{algorithm2e}

\ifCLASSINFOpdf

\else

\fi

\usepackage{amsfonts,epsfig,cite,array,multirow,graphicx,amsmath,ltablex,tabularx,setspace,amssymb,multirow}
\usepackage{schemabloc,tikz,amsthm}
\usepackage{subfig}
\usepackage[numbers,sort&compress]{natbib}
\usetikzlibrary{circuits, arrows}
\usepackage{verbatim}
\usepackage{graphicx}
\usepackage{epstopdf}
\usepackage{mathtools}
\usepackage{xcolor}
\usepackage{marvosym}

\usepackage{lipsum}
\usepackage{dblfloatfix}
\usepackage{color}
\usepackage{etoolbox}
\usepackage{pstricks}
\usepackage{epsfig}
\theoremstyle{plain}

\usepackage{tikz}
\usetikzlibrary{shapes}
\usetikzlibrary{plotmarks}
\usepackage{pgfplots}
\usetikzlibrary{spy}
\usepackage{collectbox}

\makeatletter

\DeclareMathOperator{\arccot}{arccot}

\begin{document}
\title{Performance  of Intelligent Reconfigurable Surface-Based Wireless Communications Using QAM Signaling}
\author{Dharmendra Dixit~\IEEEmembership{}, Kishor Chandra Joshi~\IEEEmembership{}, Sanjeev Sharma~\IEEEmembership{},
\thanks{ Dharmendra Dixit (e-mail: d.dixit2007@gmail.com) is with the Department of Electronics Engineering, Rajkiya Engineering College Sonbhadra, U. P. 231206, India.
Kishor Chandra (e-mail: ) is with CNRS and University of Paris-Saclay, Paris, France.
Sanjeev Sharma (e-mail: sanjeev.ece@itbhu.ac.in) is with the Department of Electronics Engineering, IIT (BHU) Varanasi .
 }}
\maketitle
\begin{abstract}
	Intelligent reconfigurable surface (IRS) is being seen as a promising technology for 6G wireless networks. The IRS can reconfigure the wireless propagation
    environment, which results in significant performance improvement 
    of wireless communications. In this paper, we analyze the performance of 
	bandwidth-efficient quadrature amplitude modulation (QAM) techniques for IRS-assisted wireless communications over Rayleigh fading channels. New closed-form expressions of the generic average symbol error rate (ASER) for rectangular QAM, square QAM  and cross QAM schemes are derived. Moreover,  simplified expressions of the ASER for low signal-to-noise-ratio (SNR) and high SNR regions are also presented, which are useful to provide insights analytically. We comprehensively analyze the impact of modulation parameters and the number of IRS elements employed. We also verify our theoretical results through simulations. Our results demonstrate that  employing IRS significantly enhances the ASER performance in comparison to additive white Gaussian noise channel at a low SNR regime.  Thus, IRS-assisted wireless communications can be a promising candidate for various low powered communication applications such as internet-of-things (IoT).

\end{abstract}


\section{Introduction}
In recent years, unprecedented growth of mobile data traffic is witnessed due to the rapid proliferation of various wireless communication  technologies, applications, and services. 
In general, 5G standardization goals are defined based on enhanced mobile broadband, ultra-reliable and low latency communications  and massive machine-type communications to address the key wireless communication requirements.
To fulfill the above requirements, different techniques such as Millimeter-wave (mmWave) communication, massive multiple-input multiple-output (MIMO), new waveform design, ultra-dense network (UDN), etc. are proposed \cite{Jefry5G}. 
Further, spectral and energy efficient wireless systems design is a key requirement for 5G and beyond wireless networks, and recently proposed  intelligent reconfigurable surface (IRS)-assisted wireless system  design can improve spectral and energy efficiency significantly by enhancing the received signal power at a user node \cite{wu2019towards}.

One common feature of  all the previous generations of wireless communications up to 5G is the assumption that  wireless channel is truly random and hence cannot be controlled. Thus the various mechanisms (e.g., MIMO, orthogonal frequency-division multiplexing (OFDM), etc.) developed for the performance improvement of wireless communication systems  over the years considered wireless channel impairments as unavoidable and tried to mitigate the adverse effects of wireless channel at the transmitter and receiver. Recently, a new technique called the IRS, which uses artificial materials to control the propagation characteristics of the radio environment,  is being seen a revolutionary technology to enhance the performance of future wireless systems~\cite{liaskos2018new,  wu2019towards, di2019smart}. Instead of compensating the destructive effects of propagation environment at receiver and transmitter, IRS allows the radio environments  to reconfigure itself in such a way that the negative effects of the environment are undone before it reaches the receiver. The ability to control wireless channel brings a true paradigm shift in the fundamentals of design and deployment of wireless communication systems since the interaction of radio signal with the radio environment can be deterministically defined to a large extent.

The key underline principle of the IRS-assisted communications is to tune the radio environment (obstacles, reflectors, etc.) such that the resulting end-to-end communication channel is favorable to radio signal transmission~\cite{basar2019transmission}. IRSs are artificial surfaces whose electromagnetic response can be electronically controllable. This allows IRS elements to reflect/refract signals at any desired angle. It becomes specifically beneficial in the context of upcoming 5G and 6G systems employing  mmWave and Terahertz bands~\cite{strinati20196g, david20186g} where providing ubiquitous coverage is very difficult due to a high signal absorption by obstacles/wireless environment.  IRSs are also important from the energy efficiency perspective as the IRSs do not involve any encoding/decoding or amplification of impinged signal while significantly improving the signal power at the intended receiver~\cite{huang2019reconfigurable}.

Recently, research on IRS-assisted communications has received tremendous attention  for wireless communications due to its potential applications in improving the coverage, energy-efficiency and data-rates . In the seminal work on IRS~\cite{wu2018intelligent}, a mathematical model for IRS-assisted communication is presented. The extension of the proposed model by employing beamforming-based design to model the IRS is presented~\cite{wu2019intelligent}. In \cite{ellingson2019path}, a detailed physical channel model considering the physical dimensions of IRS, element spacing, and radiation pattern is proposed and the relation between the size of IRS and path-loss is numerically established. In \cite{zappone2020overhead}, the impacts of overheads resulting due to feedback on the spectral efficiency are analyzed. In \cite{BasarJul2019}, authors investigate theoretical performance limits of IRS-assisted communication systems using a binary modulation scheme. In \cite{di2019reflection}, a modeling framework for IRS-assisted communications using  stochastic geometry is proposed. Performance comparison considering reflection probability defined as the availability of reflected path between transmitter and receiver with and without the IRS-coated obstacles is provided. Similarly, using the stochastic geometry approach, the application of IRS to mitigate the coverage holes in dense-urban environments is explored in \cite{kishk2020exploiting}.  In \cite{di2020analytical}, authors develop closed-form expression for reflected power from IRS using the general scalar theory of diffraction and the Hygens-Fresnel principle as a function of the transmitter/receiver  and IRS distance, phase-transformation applied by IRS and the size of IRS. The proposed approach identify the conditions in which IRS behaves as an anomalous reflector resulting in the path loss proportional to the summation of distances between the transmitter-IRS and IRS-receiver. In \cite{bjornson2019intelligent}, IRS-assisted communication is compared with the decode-and-forward relaying and relationship between the size of IRS and corresponding performance gain is derived. The $N^2$ power scaling law of IRS is revisited in \cite{bjornson2020power} and it is shown that with an increasing number of IRS elements, when the far-field assumption is no longer valid, power scaling does not change by the $N^2$ factor, where $N$ . denotes the total number of elements in the IRS.
	
Quadrature amplitude modulation~(QAM) is a generic modulation scheme, which several well-known modulation scheme as special cases. Rectangular QAM (RQAM), square QAM (SQAM) and cross QAM (XQAM) are three main types of QAM schemes.  Employing QAM has become an essential practice in modern wireless systems due to the high spectral-efficiency offered by QAMs. For example, 3GPP Release-14~(LTE-advanced) and Release-15 (5G-New radio) specifications already support up to 256-QAMs. Since IRSs can significantly  improve the signal strength without introducing any additional noise, the usage of higher order QAMs would be highly beneficial to enhance the spectral efficiency.

To the best of our knowledge, there is no work in the literature on the the performance of QAM signaling for IRS-assisted wireless communications. Motivated by applicability and importance of the aforementioned problem, a comprehensive analytical performance analysis of IRS-assisted wireless communications using QAM signals are given. In this paper, new closed-form analytical expressions for the generic average symbol error rate (ASER) of IRS-assisted wireless communications employing RQAM and XQAM signaling are derived.   Moreover, simplified expressions for both low signal-to-noise-ratio (SNR) and high SNR values are also obtained. We comprehensively analyze the different trade-offs resulting due to modulation parameters and the number of IRS elements employed. We also verify our theoretical results through simulations.   

The remaining of this paper is structured as follows. Section II presents system and channel models followed by derivations of closed form expressions in Section III. In Section IV, numerical results and discussion are presented. The conclusion of paper is given in Section V.

\section{System and Channel Models}
\begin{figure}[htb!]
	\vspace{-.1em}
	\centering
	\includegraphics[]{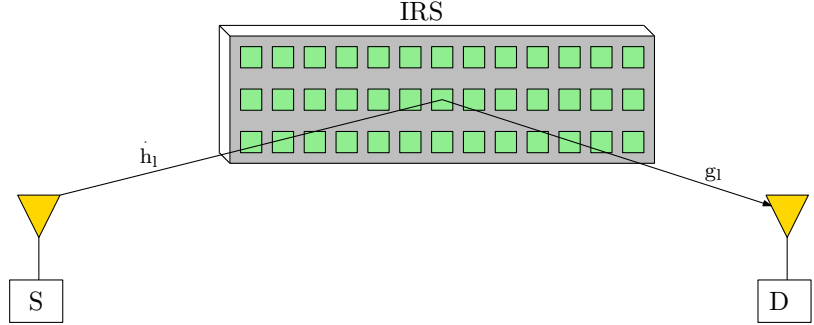}
	\caption { IRS-assisted dual-hop communication system
		without a line-of-sight path between source $S$ and destination $D$.}	 	
	\label{system}
\end{figure}

The system model consists of a source node ($S$) and  a destination node ($D$) assisted by a $N$-element IRS as shown in Fig.~\ref{system}. 
We assume that both $S$ and $D$ have a single antenna. The function of each IRS element is to introduce desired phase-shift to  incident signal. $h_l$ and $\text{g}_l$ represent the fading channel coefficients between the
$S$ and the $l$-th element of IRS, and between the $l$-th element of IRS and the $D$, respectively $(l=1,2,\ldots,N)$.  Under the assumption of Rayleigh fading channels, $h_l$ and $\text{g}_l$ can be modeled as complex Gaussian distribution with zero mean and unity variance. We assume that the $S$ transmits symbol $x$ with average energy $E_s$. Therefore, the received signal at the $D$ for a slowly varying and flat fading conditions can be written as
\begin{align}
r=\mathbf{g}^T\mathbf{\Psi}\mathbf{h}\,x+n,
\end{align}
where $\mathbf{h}=\left[|h_1|\exp\left(-j\psi_1\right),|h_2|\exp\left(-j\psi_2\right),\ldots,|h_N|\exp\left(-j\psi_N\right)\right]^T$ and\\ $\mathbf{g}=\left[|\text{g}_1|\exp\left(-j\phi_1\right),|\text{g}_2|\exp\left(-j\phi_2\right),\ldots,|\text{g}_N|\exp\left(-j\phi_N\right)\right]^T$
represent the vectors of fading coefficients between the $S$ and the IRS, and between the $D$ and the IRS, respectively, and $\mathbf{\Psi}=diag\left(\left[\exp\left(-j\varphi_1\right),\exp\left(-j\varphi_2\right),\ldots,\exp\left(-j\varphi_N\right)\right]\right)$
is a diagonal matrix that contains the phase shifts introduced 
by the $N$-element IRS. $n$ denotes the additive white Gaussian noise (AWGN) sample modeled  as complex Gaussian distribution with zero mean and $N_0$ variance. Assuming the perfect knowledge of  channel coefficients $h_l$ and $\text{g}_l$ at IRS, the instantaneous SNR at $D$ can be expressed as follows
\begin{align}
\gamma=\left(\sum_{l=1}^{N}|h_l||\text{g}_l|\right)^2\frac{E_s}{N_0}.
\end{align}
Since $|h_l|$ and $|\text{g}_l|$ are assumed to be independently Rayleigh distributed random variables (RVs), the mean value and the variance of their product are $\mathbf{E}\left[|h_l||\text{g}_l|\right]=0.25\pi$ and $\mathbf{Var}\left[|h_l||\text{g}_l|\right]=1-0.0625\pi^2$, respectively, where $\mathbf{E}\left[\cdot\right]$ is an expectation operator and $\mathbf{Var}\left[|h_l||\text{g}_l|\right]$ is a variance operator. For a sufficiently large value of $N$ i.e., $N>>1$, according to the central limit theorem, $\sum_{l=1}^{N}|h_l||\text{g}_l|$ converges to a Gaussian distributed RV with statistical parameters $\mathbf{E}\left[\sum_{l=1}^{N}|h_l||\text{g}_l|\right]=0.25\pi N$ and $\mathbf{Var}\left[\sum_{l=1}^{N}|h_l||\text{g}_l|\right]=\left(1-0.0625\pi^2\right)N$. Hence, 
 the RV $\gamma$ is a non-central chi-square RV with one degree of freedom and has the following
moment generating function (MGF) $\mathcal{G}_{\gamma}(s)$ \cite{ProakisBookDC,BasarJul2019}
\begin{align}\label{mgf}
\mathcal{G}_{\gamma}(s)&\approx \left(\frac{\Delta_1
}{\Delta_1+s\bar{\gamma}}\right)^{0.5}
\,\exp\left(-\frac{s\bar{\gamma}\Delta_2}{\Delta_1+s\bar{\gamma}}\right),
\end{align}
where $\Delta_1=\frac{8}{N(16-\pi^2)}$, $\Delta_2=\frac{N\pi^2}{2(16-\pi^2)}$, and $\bar{\gamma}=\frac{E_s}{N_0}$.
\section{Performance Analysis}
ASER is defined as the averaging of conditional symbol error rate (SER) $P_{e|\gamma}(\cdot)$
over the PDF $f_{\gamma}(\cdot)$ of SNR $\gamma$ \cite{SimonBookDC}, i.e.,
\begin{equation}\label{ASER}
P_{e}= \mathbb{E}\left[ P_{e|\gamma}(x)\right]=\int_{0}^{\infty} P_{e|\gamma}(x)f_{\gamma}(x)dx.
\end{equation}
\subsection{$M$-ary Rectangular QAM}
The expression of conditional SER $P_{e|\gamma}(x)$ for $M$-ary RQAM is given as \cite{BeaulieuMay2006}
\begin{align}\label{CSERRQAM1}
 P_{e|\gamma}(x)&=2p\,Q\left(a\sqrt{x}\right)+2q\,Q\left(b\sqrt{x}\right)
-4p\,q\,Q\left(a\sqrt{x}\right) Q\left(b\sqrt{x}\right),
\end{align}
where $M=M_I\times M_Q$, $p=1-\frac{1}{M_I}$, $q=1-\frac{1}{M_Q}$,
$a=\sqrt{\frac{6}{(M_I^2-1)+(M_Q^2-1)\beta^2}}$, $b=\beta a$ and $\beta=d_Q/d_I$ is the quadrature-to-in-phase decision distance ratio with $d_I$ and $d_Q$ being
the in-phase and quadrature decision distance, respectively and $Q(t)$ is the Gaussian $Q$-function defined by
$ Q(t)=\frac{1}{\sqrt{2\pi}}\int_{t}^{\infty}\mbox{exp}(-\frac{x^2}{2})dx$.
For the convenience of analysis, (\ref{CSERRQAM1}) can be rewritten using an alternate expression of $1$-$D$ and $2$-$D$ Gaussian $Q$-functions, $Q_z(t,\phi)$ defined as \cite{SimonBookDC}
 \begin{eqnarray}\label{qfunction1}
 Q_z(t,\theta)=\frac{1}{\pi}\int_{0}^{\theta}\exp\left(-\frac{t^2}{2\sin^2\phi}\right)d\phi,\qquad t\geq 0.
 \end{eqnarray}
With the aid of (\ref{qfunction1}), it can be obtained that the  $Q(t)=Q_z\left(t,\pi/2\right)$ \cite[(4.2)]{SimonBookDC} and $Q(t)Q(u)=0.5\big[Q_z\left(u,\arctan(u/t)\right)
+Q_z\left(t,\arccot(u/t)\right)\big]$ \cite[(4.8)]{SimonBookDC} for $t\geq 0,u\geq 0$.
Thus, employing (\ref{qfunction1}), (\ref{CSERRQAM1}) has been expressed as
 \begin{align}\label{CSERRQAM2}
 P_{e|\gamma}(x)&=2p\,Q_z\left(a\sqrt{x},\pi/2\right)
 +2q\,Q_z\left(b\sqrt{x},\pi/2\right)
-2p\,q\,\Big[Q_z\left(b\sqrt{x},\arctan(b/a)\right)
\nonumber \\ &
+Q_z\left(a\sqrt{x},\arccot(b/a)\right) 
\Big].
\end{align}
Substituting (\ref{CSERRQAM2}) into (\ref{ASER}) and by doing a few simplifications,  ASER for RQAM scheme denoted as $ P_e^{\text{RQAM}}$, can be written as
\begin{align}\label{ASERRQAM}
 P_e^{\text{RQAM}}
&=2p\,\mathcal{I}\left(a,\pi/2\right)+2q\,\mathcal{I}\left(b,\pi/2\right)
-2p\,q\Big[\mathcal{I}\left(b,\arctan(b/a)\right)
+\mathcal{I}\left(a,\arccot(b/a)\right)
\Big],
\end{align}
where the integral $\mathcal{I}\big(\cdot,\cdot)$ is expressed as
\begin{align}\label{I}
\mathcal{I}\left(c,\theta\right)&=\int_{0}^{\infty}Q_z(c\sqrt{x},\theta)f_\gamma(x)\, dx
\nonumber \\ & =\frac{1}{\pi}\int_0^\theta \int_{0}^{\infty}\mbox{exp}\Big(-\frac{c^2 x}{2\sin^2\phi}\Big) f_\gamma(x)\, dx\, d\phi
\nonumber\\ &=\frac{1}{\pi}\int_0^\theta
\mathcal{G}_{\gamma}\Big(\frac{c^2}{2\sin^2\phi}\Big)d\phi.
\end{align}
Here $\mathcal{G}_{\gamma}\left(\cdot\right)$ is the MGF of $\gamma$. Thus, to obtain a solution
of (\ref{I}), we need  the MGF of
$\gamma$ which is given in (\ref{mgf}). 
Substituting (\ref{mgf}) in (\ref{I}), closed-form solutions for
$\mathcal{I}(x, \pi/2)$, $\mathcal{I}(b, \arctan(b/a))$ and
$\mathcal{I}(x,\arccot(b/a))$ , which will be employed to compute (\ref{ASERRQAM}), are obtained in (\ref{Iapiby2}), (\ref{akarctanba}) and (\ref{Iapibyarctan}), respectively,
in Appendix~\ref{AppendixA}. Thus, substituting these solutions in (\ref{ASERRQAM}) and simplifying the resulting expression, an ASER expression for $P_e^{\text{RQAM}}$ can be obtained as in (\ref{ASERRQAM_Final}).
\begin{figure*}[t!]
\normalsize
\setcounter{equation}{9}
\begin{align}\label{ASERRQAM_Final}
P_e^{\text{RQAM}}
&\approx\frac{2 p a\sqrt{2\Delta_1\bar{\gamma}}\exp\left(-\Delta_2\right)}{\pi (a^2\bar{\gamma}+2\Delta_1)}\Phi_1^{(2)}\left(1;1;1.5; \frac{2\Delta_1}{a^2\bar{\gamma}+2\Delta_1},
\frac{2\Delta_1\Delta_2}{a^2\bar{\gamma}+2\Delta_1}\right)
\nonumber \\ &
+\frac{2 q b\sqrt{2\Delta_1\bar{\gamma}}\exp\left(-\Delta_2\right)}{\pi (b^2\bar{\gamma}+2\Delta_1)}\Phi_1^{(2)}\left(1;1;1.5; \frac{2\Delta_1}{b^2\bar{\gamma}+2\Delta_1},
\frac{2\Delta_1\Delta_2}{b^2\bar{\gamma}+2\Delta_1}\right)
\nonumber \\ &
-\frac{2p q b\sqrt{2\Delta_1\bar{\gamma}}\exp\left(-\Delta_2\right)}{2\pi ((a^2+b^2)\bar{\gamma}+2\Delta_1)}
\Phi_1^{(3)}\left(1;0.5,1;2; \frac{(b^2\,\bar{\gamma}+2\Delta_1)}{(a^2+b^2)\bar{\gamma}+2\Delta_1},
\right.
\nonumber\\&\left.
\frac{2\Delta_{1}}{(a^2+b^2)\bar{\gamma}+2\Delta_1},
\frac{2\Delta_1\Delta_{2}}{(a^2+b^2)\bar{\gamma}+2\Delta_1}
\right)
-\frac{2p q a\sqrt{2\Delta_1\bar{\gamma}}\exp\left(-\Delta_2\right)}{2\pi ((a^2+b^2)\bar{\gamma}+2\Delta_1)}
\nonumber \\ & \times
\Phi_1^{(3)}\left(1;0.5,1;2; \frac{a^2\,\bar{\gamma}+2\Delta_1}{(a^2+b^2)\bar{\gamma}+2\Delta_1},
\frac{2\Delta_{1}}{(a^2+b^2)\bar{\gamma}+2\Delta_1},
\frac{2\Delta_1\Delta_{2}}{(a^2+b^2)\bar{\gamma}+2\Delta_1}
\right).
\end{align}
\hrulefill
\end{figure*}
For the special case of $M$-ary SQAM, i.e.  when $M_I = M_Q = \sqrt{M}$ and $\beta = 1$, one can get ASER expression of $M$-ary SQAM which is provided in (\ref{ASERSQAM_Final}),
\begin{figure*}[t!]
\normalsize
\setcounter{equation}{10}
\begin{align}\label{ASERSQAM_Final}
P_e^{\text{SQAM}}
&\approx\frac{4 \tilde{p} \tilde{a}\sqrt{2\Delta_1\bar{\gamma}}\exp\left(-\Delta_2\right)}{\pi (\tilde{a}^2\bar{\gamma}+2\Delta_1)}\Phi_1^{(2)}\left(1;1;1.5; \frac{2\Delta_1}{\tilde{a}^2\bar{\gamma}+2\Delta_1},
\frac{2\Delta_1\Delta_2}{\tilde{a}^2\bar{\gamma}+2\Delta_1}\right)
\nonumber \\ &
-\frac{\tilde{p}^2 \tilde{a}\sqrt{2\Delta_1\bar{\gamma}}\exp\left(-\Delta_2\right)}{\pi (\tilde{a}^2\bar{\gamma}+\Delta_1)}
\Phi_1^{(3)}\left(1;0.5,1;2; \frac{(\tilde{a}^2\,\bar{\gamma}+2\Delta_1)}{2\tilde{a}^2\bar{\gamma}+2\Delta_1},
\frac{\Delta_{1}}{\tilde{a}^2\bar{\gamma}+\Delta_1},
\frac{\Delta_1\Delta_{2}}{\tilde{a}^2\bar{\gamma}+\Delta_1}
\right)
\end{align}
\hrulefill
\end{figure*}
where $\tilde{p}=1-\frac{1}{\sqrt{M}}$ and $\tilde{a}=\sqrt{\frac{3}{M-1}}$.
Similarly, for BPSK, i.e.  when $M_I = 2, M_Q = 1$ and $\beta = 0$, one can obtain ASER expression of BPSK which is given in (\ref{ASERBPSK_Final}). It is important to mention that (\ref{ASERSQAM_Final}) and (\ref{ASERBPSK_Final}) are the closed-form solutions of \cite[(21)]{BasarJul2019} and \cite[(17)]{BasarJul2019}, respectively.

\begin{figure*}[t!]
\normalsize
\setcounter{equation}{11}
\begin{align}\label{ASERBPSK_Final}
P_e^{\text{BPSK}}
&\approx\frac{ \sqrt{\Delta_1\bar{\gamma}}\exp\left(-\Delta_2\right)}{\pi (\bar{\gamma}+\Delta_1)}\Phi_1^{(2)}\left(1;1;1.5; \frac{\Delta_1}{\bar{\gamma}+\Delta_1},
\frac{\Delta_1\Delta_2}{\bar{\gamma}+\Delta_1}\right).
\end{align}
\hrulefill
\end{figure*}

\subsection{$M$-ary Cross QAM}
The conditional SER $P_{e|\gamma}(x)$ for $M$-ary XQAM can be given as \cite{YuJul2011}
\begin{align}\label{CSERXQAM}
P_{e|\gamma}(x)&=w_1 Q_z\left(a_0\sqrt{x},\pi/2\right)
+w_2 Q_z\left(a_1\sqrt{x},\pi/2\right)
-w_3 Q_z\left(a_0\sqrt{x},\pi/4\right)
\nonumber\\&
-2 w_2 \sum_{l=1}^{L-1}Q_z\left(a_0\sqrt{x},\alpha_l\right)
-w_2\sum_{l=1}^{L-1}Q_z\left(a_l\sqrt{x},\beta_l^+\right)
+w_2\sum_{l=2}^{L}Q_z(a_l\sqrt{x},\beta_l^-),
\end{align}
where $w_1=4-\frac{6}{\sqrt{2M}}$, $w_2=\frac{4}{M}$, $w_3=4-\frac{12}{\sqrt{2M}}+\frac{12}{M}$, $L=\frac{\sqrt{2M}}{8}$, $a_0=\sqrt{\frac{96}{(31 M-32)}}$,
$a_l=\sqrt{2} l a_0,\, l=1,2, \ldots, L$, $\alpha_l=\arctan\left(\frac{1}{2l+1}\right),\, l=1,2, \ldots, (L-1)$;
$\beta_l^-=\arctan\left(\frac{l}{l-1}\right),\, l=2, 3, \ldots, L$;
$\beta_l^+=\arctan\left(\frac{l}{l+1}\right),\, l=1, 2, \ldots, L-1$.

Substituting (\ref{CSERXQAM}) into (\ref{ASER}) and using (\ref{qfunction1}), the ASER expression of XQAM can be written as
\begin{align}\label{ASERXQAM}
P_e^{\text{XQAM}}
 &=w_1\,\mathcal{I}\left(a_0,\pi/2\right)+w_2\,\mathcal{I}\left(a_1,\pi/2\right)
-w_3\,\mathcal{I}\left(a_0,\pi/4\right)-2 w_2\sum_{l=1}^{L-1}\mathcal{I}\left(a_0,\alpha_l\right)
\nonumber \\ & 
-w_2\sum_{l=1}^{L-1}\mathcal{I}\left(a_l,\beta_l^+\right)
+w_2\sum_{l=2}^{L}\mathcal{I}\left(a_l,\beta_l^-\right),
\end{align}
where $\mathcal{I}\big(\cdot,\cdot)$ is given in (\ref{I}).
To obtain a closed-form solution of (\ref{ASERXQAM}),
it is required to obtain the solution of $\mathcal{I}(c,\theta)$. The desired solutions are given in Appendix~\ref{AppendixA}. Employing a similar method as followed to obtain (\ref{ASERRQAM_Final}), an ASER expression of XQAM can be written as (\ref{ASERXQAM_Final}), at the top of the next page.
 \begin{figure*}[t!]
\normalsize
\setcounter{equation}{14}
\begin{align}
 P_e^{\text{XQAM}}
 &\approx 
 \frac{w_1 a_0\sqrt{2\Delta_1\bar{\gamma}}\exp\left(-\Delta_2\right)}{\pi (a_0^2\bar{\gamma}+2\Delta_1)}\Phi_1^{(2)}\left(1;1;1.5; \frac{2\Delta_1}{a_0^2\bar{\gamma}+2\Delta_1},
\frac{2\Delta_1\Delta_2}{a_0^2\bar{\gamma}+2\Delta_1}\right)
 \nonumber \\ & 
 +\frac{w_2 a_1\sqrt{2\Delta_1\bar{\gamma}}\exp\left(-\Delta_2\right)}{\pi (a_1^2\bar{\gamma}+2\Delta_1)}\Phi_1^{(2)}\left(1;1;1.5; \frac{2\Delta_1}{a_1^2\bar{\gamma}+2\Delta_1},
\frac{2\Delta_1\Delta_2}{a_1^2\bar{\gamma}+2\Delta_1}\right)
\nonumber \\ & 
-\frac{w_3 a_0\sqrt{2\Delta_1\bar{\gamma}}\exp\left(-\Delta_2\right)}{4\pi (a_0^2\bar{\gamma}+\Delta_1)}
\Phi_1^{(3)}\left(1;0.5,1;2; \frac{(a_0^2\,\bar{\gamma}+2\Delta_1)}{2a_0^2\bar{\gamma}+2\Delta_1},
\frac{\Delta_{1}}{a_0^2\bar{\gamma}+\Delta_1},
\frac{\Delta_1\Delta_{2}}{a_0^2\bar{\gamma}+\Delta_1}
\right)
\nonumber \\ & 
-2 w_2\sum_{l=1}^{L-1}
\frac{a_0\,\sqrt{2\Delta_1\bar{\gamma}}\exp\left(-\Delta_2\right)}{2\pi (a_0^2(1+(2l+1)^2)\bar{\gamma}+2\Delta_1)}
\Phi_1^{(3)}\left(1;0.5,1;2; \frac{(a_0^2\,\bar{\gamma}+2\Delta_1)}{a_0^2(1+(2l+1)^2)\bar{\gamma}+2\Delta_1},
\right.
\nonumber\\&\left.
\frac{2\Delta_{1}}{a_0^2(1+(2l+1)^2)\bar{\gamma}+2\Delta_1},
\frac{2\Delta_1\Delta_{2}}{a_0^2(1+(2l+1)^2)\bar{\gamma}+2\Delta_1}
\right)
\nonumber \\ & 
-w_2\sum_{l=1}^{L-1}
\frac{a_l\,l^2\sqrt{2\Delta_1\bar{\gamma}}\exp\left(-\Delta_2\right)}{2\pi (a_l^2(l^2+(l-1)^2)\bar{\gamma}+2\Delta_1\,l^2)}
\Phi_1^{(3)}\left(1;0.5,1;2; \frac{(a_l^2\,\bar{\gamma}+2\Delta_1)l^2}{a_l^2(l^2+(l-1)^2)\bar{\gamma}+2\Delta_1\,l^2},
\right.
\nonumber\\&\left.
\frac{2\Delta_{1}\,l^2}{a_l^2(l^2+(l-1)^2)\bar{\gamma}+2\Delta_1\,l^2},
\frac{2\Delta_1\Delta_{2}\,l^2}{a_l^2(l^2+(l-1)^2)\bar{\gamma}+2\Delta_1\,l^2}
\right)
\nonumber \\ & 
+w_2\sum_{l=2}^{L}
\frac{a_l\,l^2\sqrt{2\Delta_1\bar{\gamma}}\exp\left(-\Delta_2\right)}{2\pi (a_l^2(l^2+(l+1)^2)\bar{\gamma}+2\Delta_1\,l^2)}
\Phi_1^{(3)}\left(1;0.5,1;2; \frac{(a_l^2\,\bar{\gamma}+2\Delta_1)l^2}{a_l^2(l^2+(l+1)^2)\bar{\gamma}+2\Delta_1\,l^2},
\right.
\nonumber\\&\left.
\frac{2\Delta_{1}\,l^2}{a_l^2(l^2+(l+1)^2)\bar{\gamma}+2\Delta_1\,l^2},
\frac{2\Delta_1\Delta_{2}\,l^2}{a_l^2(l^2+(l+1)^2)\bar{\gamma}+2\Delta_1\,l^2}
\right)
\label{ASERXQAM_Final}  
\end{align}
\hrulefill
\end{figure*}
\subsection{Low SNR Analysis for IRS}
At low SNR region, one can approximate MGF expression in (\ref{mgf}) as given below
\begin{align}\label{mgflowsnr}
\mathcal{G}_{\gamma}^{L}(s)&\approx
\exp\left(-\frac{\Delta_2}{\Delta_1}s\bar{\gamma}\right),
\end{align}
Hence, the ASER expression of RQAM at low SNR region can be expressed as
\begin{align}\label{ASERRQAMlowsnr}
P_{e,\text{low SNR}}^{\text{RQAM}}
&=2p\,\mathcal{I}^L\left(a,\pi/2\right)+2q\,\mathcal{I}^L\left(b,\pi/2\right)
-2p\,q\Big[\mathcal{I}^L\left(b,\arctan(b/a)\right)
+\mathcal{I}^L\left(a,\arccot(b/a)\right)
\Big],
\end{align}
where the integral $\mathcal{I}^L\big(\cdot,\cdot)$ is given as
\begin{align}\label{Ilow}
\mathcal{I}^L\left(c,\theta\right)&=\frac{1}{\pi}\int_0^\theta
\mathcal{G}_{\gamma}^L\left(\frac{c^2}{2\sin^2\phi}\right)d\phi,
\end{align}
In order to get insights, (\ref{Ilow}) can be upper bounded
by setting $\phi=\theta$, which gives
\begin{align}\label{Ilow2}
\mathcal{I}^L\left(c,\theta\right)&\leq\frac{\theta}{\pi}
\mathcal{G}_{\gamma}^L\left(\frac{c^2}{2\sin^2\theta}\right)\nonumber\\
&\leq\frac{\theta}{\pi}\exp\left(-\frac{\Delta_2\,c^2\,\bar{\gamma}}{2\Delta_1\,\sin^2\theta}\right)
\nonumber\\
&\leq\frac{\theta}{\pi}\exp\left(-\frac{N^2\pi^2\,c^2\,\bar{\gamma}}{32\,\sin^2\theta}\right)
\end{align}
Using (\ref{Ilow2}), we can write the ASER expression of RQAM at low SNR region 
\begin{align}\label{ASERRQAMlowsnrFinal}
P_{e,\text{low SNR}}^{\text{RQAM}}
&=p\,\exp\left(-\left(\frac{\pi^2\,a^2}{32}\right)N^2\,\bar{\gamma}\right)
+q\,\exp\left(-\left(\frac{\pi^2\,b^2}{32}\right)N^2\,\bar{\gamma}\right)
\nonumber \\ &
-\frac{2p\,q}{\pi}\Bigg[\arctan\left(\frac{b}{a}\right)
\exp\left(-\left(\frac{\pi^2\,(a^2+b^2)}{32}\right)N^2\,\bar{\gamma}\right)
\nonumber \\ & 
+\arccot\left(\frac{b}{a}\right)\exp\left(-\left(\frac{\pi^2\,(a^2+b^2)}{32}\right)N^2\,\bar{\gamma}\right)
\Bigg].
\end{align}
Therefore, using (\ref{ASERRQAMlowsnrFinal}) we can also state that
\begin{align}\label{ASERRQAMlowsnrFinal2}
P_{e,\text{low SNR}}^{\text{RQAM}}
&\propto\eta_1 \exp\left(-\eta_2\, N^2\,\bar{\gamma}\right)
\end{align}
where $\eta_1$ and $\eta_2$ depend on modulation order $M$. Similarly, we can also obtain the ASER expression of XQAM at low SNR region.  
\subsection{High SNR Analysis for IRS}
At high SNR region, one can approximate MGF expression in (\ref{mgf}) as given below
\begin{align}\label{mgfhighsnr}
\mathcal{G}_{\gamma}^{H}(s)&\approx\left(\frac{\Delta_1}{s\bar{\gamma}}\right)^{0.5}
\exp\left(-\Delta_2\right).
\end{align}
Hence, the ASER expression of RQAM at high SNR region can be expressed as
\begin{align}\label{ASERRQAMhighsnr}
P_{e,\text{high SNR}}^{\text{RQAM}}
&=2p\,\mathcal{I}^H\left(a,\pi/2\right)+2q\,\mathcal{I}^H\left(b,\pi/2\right)
-2p\,q\Big[\mathcal{I}^H\left(b,\arctan(b/a)\right)
+\mathcal{I}^H\left(a,\arccot(b/a)\right)
\Big],
\end{align}
where the integral $\mathcal{I}^H\big(\cdot,\cdot)$ is given as
\begin{align}\label{Ihigh}
\mathcal{I}^H\left(c,\theta\right)&=\frac{1}{\pi}\int_0^\theta
\mathcal{G}_{\gamma}^H\left(\frac{c^2}{2\sin^2\phi}\right)d\phi,
\end{align}
In order to obtain insights, a closed-form solution of (\ref{Ihigh}) is needed, which is given in Appendix~\ref{AppendixB}. 
Using (\ref{IHapiby2}), (\ref{IHarctan}), and (\ref{IHarccot}), we can write the ASER expression of RQAM at high SNR region 
\begin{align}\label{ASERRQAMhighsnrFinal}
P_{e,\text{high SNR}}^{\text{RQAM}}
&=\frac{2p\,\sqrt{2\Delta_1}\exp\left(-\Delta_2\right)}{\pi a\sqrt{\bar{\gamma}}}
+\frac{2q\sqrt{2\Delta_1}\exp\left(-\Delta_2\right)}{\pi b\sqrt{\bar{\gamma}}}
-2p\,q\Bigg[
\frac{\sqrt{2\Delta_1}\exp\left(-\Delta_2\right)}{\pi b\sqrt{\bar{\gamma}}}
\left(1-\frac{a}{\sqrt{2}b}\right)
\nonumber \\ & 
+ 
\frac{\sqrt{2\Delta_1}\exp\left(-\Delta_2\right)}{\pi a\sqrt{\bar{\gamma}}}
\left(1-\frac{a}{\sqrt{a^2+b^2}}\right)
\Bigg].
\end{align}
Therefore, using (\ref{ASERRQAMhighsnrFinal}) we can also state that
\begin{align}\label{ASERRQAMhighsnrFinal2}
P_{e,\text{low SNR}}^{\text{RQAM}}
&\propto \mu \,\sqrt{\frac{\Delta_1}{\bar{\gamma}}}\exp\left(-\Delta_2\right),
\end{align}
where $\mu$ depends on modulation order $M$. Similarly, we can also obtain the ASER expression of XQAM at low SNR region.
Likewise, one can also write the ASER expression of XQAM at high SNR region. 
\subsection{Upper Bound of ASER of QAM in AWGN Channel} 
We can obtain upper bound of $Q_z(x,\theta)$, i.e.,
 \begin{align}\label{qfunction2}
Q_z(x,\theta)&=\frac{1}{\pi}\int_{0}^{\theta}\exp\left(-\frac{x^2}{2\sin^2\phi}\right)d\phi
\nonumber\\&\leq \frac{\theta}{\pi}\exp\left(-\frac{x^2}{2\sin^2\theta}\right)
\end{align}
Using (\ref{qfunction2}), we can get the upper bound ASER expression of RQAM in AWGN channel \cite{BeaulieuMay2006}
\begin{align}\label{ASERRQAMAWGN}
P_{e}^{\text{RQAM}}&\leq p\exp\left(-\frac{a^2\bar{\gamma}}{2}\right)
+q\exp\left(-\frac{b^2\bar{\gamma}}{2}\right)
-\frac{2p\,q}{\pi}\,\Big[ \arctan\left(\frac{b}{a}\right)
+\arccot\left(\frac{b}{a}\right)
\Big]
\nonumber \\ & \times
\exp\left(-\frac{(a^2+b^2)\bar{\gamma}}{2}\right).
\end{align}
Using (\ref{ASERRQAMAWGN}), we can state that
\begin{align}\label{ASERRQAMAWGN2}
P_{e}^{\text{RQAM}}&\propto\kappa_1
\exp\left(-\kappa_2\,\bar{\gamma}\right).
\end{align}
where $\kappa_1$ and $\kappa_2$ depend on modulation order $M$. Using $\exp(-x)\approx1-x$ for small values of $x$, we can obtain
\begin{align}\label{ASERRQAMAWGNlowsnr}
P_{e,\text{low SNR}}^{\text{RQAM}}&\propto\kappa_1
\left(1-\kappa_2\,\bar{\gamma}\right).
\end{align}
\section{Numerical Results and Discussion}
We present numerical results of the ASER for various QAM signaling. We plot the performance curves of the ASER $P_e$ versus the SNR, $\bar{\gamma}$ (in dB). Simulation results are compared with numerical results for verification. 

\begin{figure}[t!]
	\begin{center}
		\includegraphics[width=12cm,height=9cm]{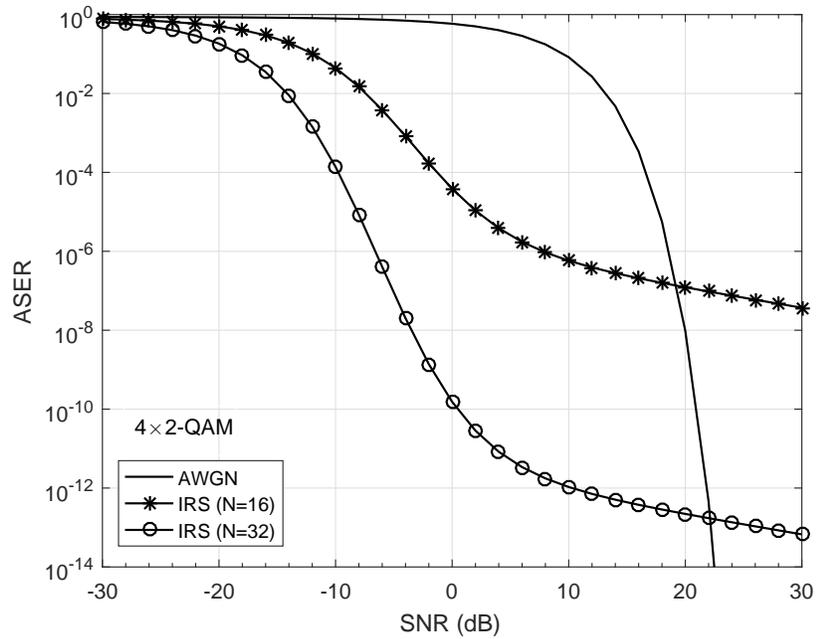}
		\caption{ASER performance comparison of IRS-based scheme for $N=16$ and $N=32$ and AWGN with $4\times2$-QAM.}
		\label{f2}
	\end{center}
\end{figure}

Fig.~\ref{f2} depicts ASER performance of $4 \times 2$-QAM for both AWGN and IRS-assisted wireless communications. It can be observed from  Fig.~\ref{f2} that an IRS-assisted scheme perform better in terms of ASER compared with an AWGN channel on low SNR regime. This is due to the fact that ASER, $P_e$ for IRS-assisted scheme at low SNR is proportional to $\eta_1\exp\left(-\eta_2 N^2\bar{\gamma}\right)$, while, ASER, $P_e$ for AWGN at low SNR is proportional to $\kappa_1\left(1-\kappa_1\bar{\gamma}\right)$. On the other hand, ASER, $P_e$ for IRS-assisted scheme at high SNR is slowly-decaying in comparison to AWGN scheme. This is due to the fact that  ASER, $P_e$ for IRS-assisted scheme at high SNR is proportional to $\mu \exp\left(-N\right)/\sqrt{N\bar{\gamma}}$, while, ASER, $P_e$ for AWGN at high SNR is proportional to $\kappa_1\exp\left(-\kappa_1\bar{\gamma}\right)$. Further, IRS-assisted system can achieve  small probability of error even very low SNR region (below $0$ dB) using medium  and large values of $N$, which can be used for various low powered IoT applications.

\begin{figure}[t!]
\begin{center}
\includegraphics[width=12cm,height=9cm]{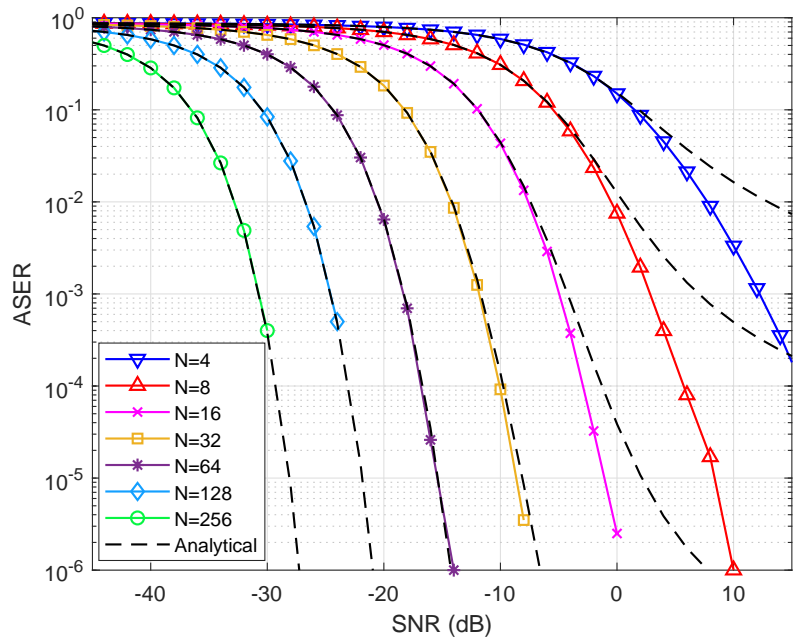}
\caption{ASER performance of IRS-assisted scheme for $4\times 2$-QAM signaling with varying $N$.}
\label{f3}
\end{center}
\end{figure}

In Fig.~\ref{f3}, we plot the ASER curves of an IRS-assisted 
scheme for different numbers of intelligent refectling surfaces $N$
and for $4\times2$-QAM signaling. Fig.\ref{f3} verifies that our
analytical approximation in (\ref{mgf}) that is based on the central limit theorem
is sufficiently accurate for large values of $N$. For smaller values of $N$ ($N=4,8, 16$), both analytical and simulation results do not match at high SNR region due to poor accuracy of approximation. However, for medium ($N=32, 63$) and large  $N=128, 256$ values of $N$, both simulation and analytical results are matched closely for the whole SNR region. 

\begin{figure}[t!]
\begin{center}
\includegraphics[width=12cm,height=9cm]{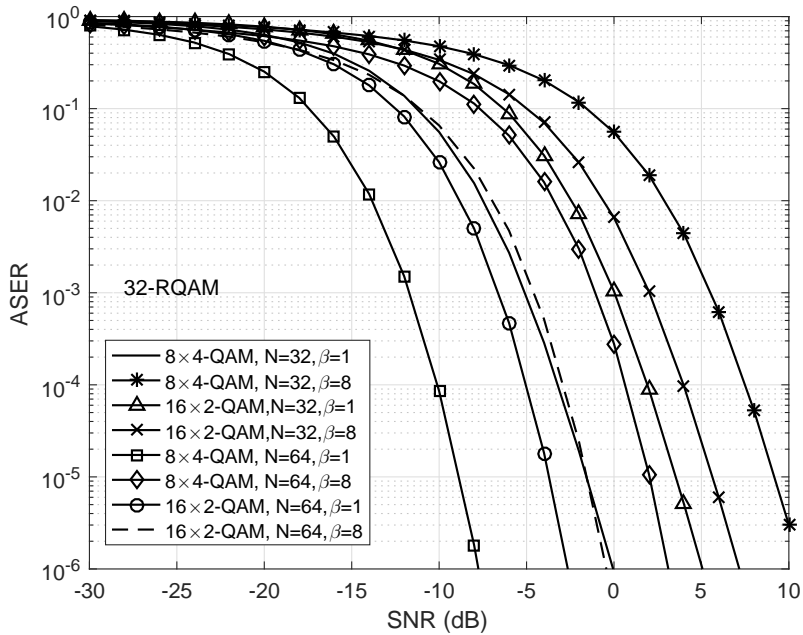}
\caption{ ASER performance of IRS- assisted scheme for $8\times 4$ and $16\times 2$ constellations with varying $N$ and $\beta$.}
\label{f4}
\end{center}
\end{figure}

\begin{figure}[t!]
\begin{center}
\includegraphics[width=12cm,height=9cm]{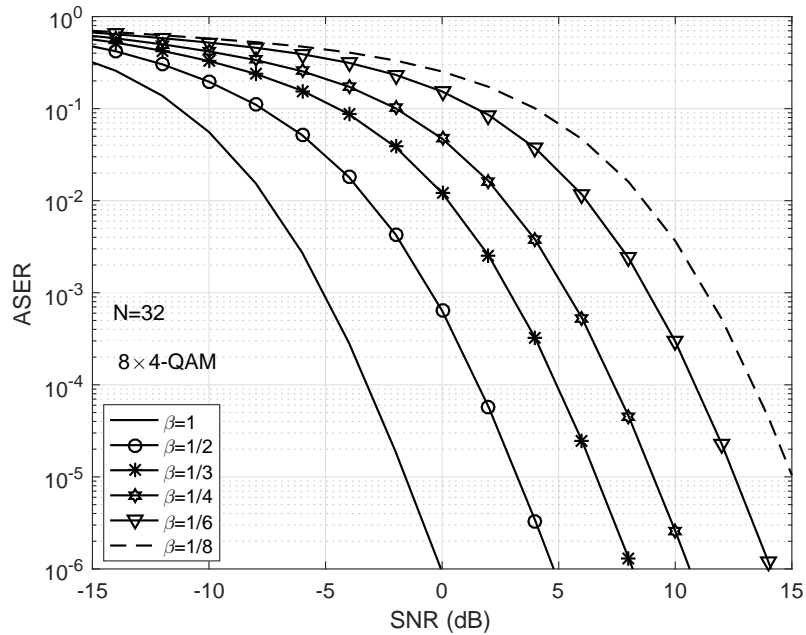}
\caption{ASER performance of IRS- assisted scheme for $32$-RQAM ($8\times 4$) signaling with $\beta\leq1$.}
\label{f5}
\end{center}
\end{figure}

Fig.~\ref{f4} shows the impact of $N$ and $\beta$ on ASER for two different constellations of a $32$-QAM scheme such as $8\times 4$ and $16\times 2$.  We have chosen only two values of $\beta$ ($\beta=1$ and $\beta=8$) for clarity in the figure. It turns out that the ASER performance improves with an increase in $N$, as expected. For a fixed $N$,  the ASER performance can be controlled by the modulation parameter $\beta$. For $8\times 4$-QAM, a minimum ASER achieves for $\beta=1$, for any value of $N$. Another observation, with increase in $\beta$, ASER for the $8\times 4$ constellation degrades fast in comparison $16\times2$ constellation. This can be explained from the relation among $M_I$, $M_Q$, and $\beta$.
Considering a range of values of $M_I$, $M_Q$, and $\beta$, we observed that the ASER of a QAM for a fixed size (i.e., $M$) may vary with different constellations. For a fixed $\beta$, the ASER can be controlled by $|M_I-M_Q|$; constellation with a large $|M_I-M_Q|$ causes poor ASER relative to the constellations with a small difference. It is understood with the definition of $\beta$ such as more number of constellation points in a quadrant brings them closer resulting in poor ASER. Further, for a QAM of for a fixed $M$, a constellation with smallest $|M_I-M_Q|$ and $\beta=1$ is better than other constellation arrangements as it achieves a minimum ASER. In $32$-QAM, the $8\times 4$ constellation with $\beta = 1$ outperforms the $16\times 2$ constellation with any $\beta$.
Fig.~\ref{f5} includes ASER curves of $8\times 4$-QAM with $\beta\leq 1$ for illustration and comparison.


\begin{figure}[t!]
\begin{center}
\includegraphics[width=12cm,height=9cm]{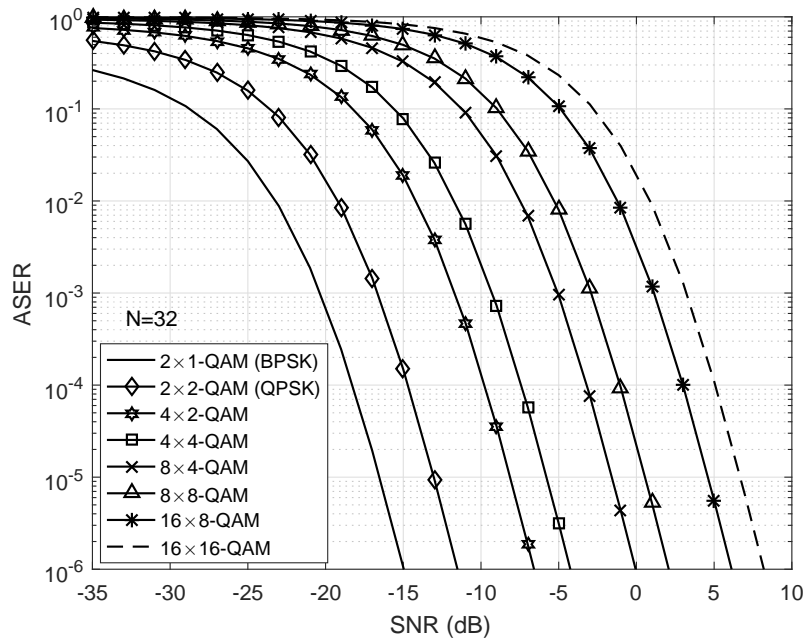}
\caption{Comparison of ASER performance of various RQAM for fixed $N=32$.}
\label{f6}
\end{center}
\end{figure}

\begin{figure}[t!]
\begin{center}
\includegraphics[width=12cm,height=9cm]{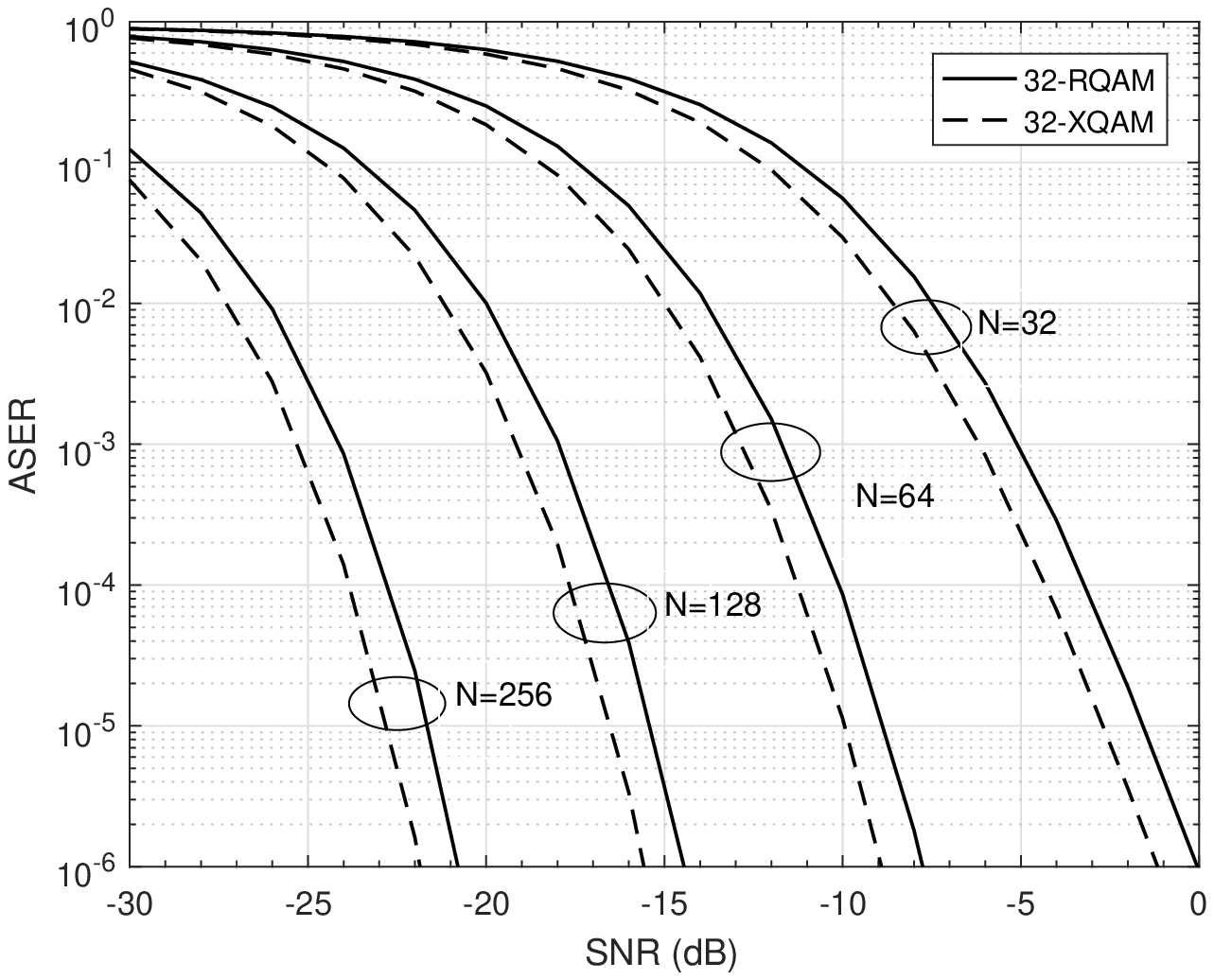}
\caption{Comparison of ASER performance of $32$-RQAM and $32$-XQAM for varying $N$.}
\label{f7}
\end{center}
\end{figure}

Fig.~\ref{f6} shows the impact of modulation order $M$ on ASER performance of IRS-based wireless communications. In comparison, we have chosen 7 different RQAM from $M=2$ to $M=256$.
It can be noticed from Fig.~\ref{f6} that for a fixed ASER, the increase in $M$ requires high SNR, as expected. 
Fig.~\ref{f7} illustrates the comparison between RQAM and
XQAM with varying $N$. For the purpose of comparing we have chosen $32$-RQAM ($8\times 4$, and $\beta=1$) and $32$-XQAM. It can be observed that XQAM achieves
SNR gain over RQAM for a given ASER. Further, the high SNR gain can be achieved by increasing the value of $N$. However,  the SNR gain between XQAM and RQAM for a given ASER does not change by varying $N$.

\section{Conclusion}
We studied the ASER performance of IRS-based radio communications  over Rayleigh fading channels. New closed-form expressions of ASER for generic SNR, low SNR and high SNR regions for the considered system employing different types of QAMs signaling were obtained. The obtained analytic expressions were used to analyze the impact of different modulation parameters: $M$ and $\beta$, and the number of IRS elements, $N$ on the system performance. It was observed that an IRS-assisted scheme outperforms AWGN scheme on low SNR regime. This is due to the fact that ASER, $P_e$ for IRS-assisted scheme at low SNR is proportional to $\eta_1\exp\left(-\eta_2 N^2\bar{\gamma}\right)$, while, ASER, $P_e$ for AWGN at low SNR is proportional to $\kappa_1\left(1-\kappa_1\bar{\gamma}\right)$. While, ASER, $P_e$ for IRS-assisted scheme at high SNR is slowly-decaying in comparison to AWGN scheme. This is due to the fact that  ASER, $P_e$ for IRS-assisted scheme at high SNR is proportional to $\mu \exp\left(-N\right)/\sqrt{N\bar{\gamma}}$, while, ASER, $P_e$ for AWGN at high SNR is proportional to $\kappa_1\exp\left(-\kappa_1\bar{\gamma}\right)$.

\appendices
\section{Solution of $\mathcal{I}\left(x,\theta\right)$}
\label{AppendixA}	
The closed-form solutions for $\mathcal{I}\left(x,\theta\right)$ are derived in terms of the  confluent Lauricella's hypergeometric function, which is defined as \cite{ExtonBook,MartinezSep2014}
	\begin{align}
	\label{phi13}
\Phi_1^{(n)}(m;p_1, p_2\ldots p_{n-1};q;z_1,z_2,\ldots z_n)
	&=\frac{\Gamma(q)}{\Gamma(m)\Gamma(q-m)}
	\int_0^1v^{m-1}
	(1-v)^{q-m-1}
	\nonumber \\& \times
	\prod_{i=1}^{n-1}(1-v\,z_i)^{-p_i}\,\exp(v\,z_n)dv,
	\end{align}
where $\Gamma(\cdot)$ denotes the Gamma function. This confluent hypergeometric function can be numerically computed with the aid of its finite integral expression.
\subsection{Closed-form Solution of $\mathcal{I}\left(x,\pi/2\right)$}
\label{AppendixAA}
 Substituting, (\ref{mgf}) into (\ref{I}) followed by
$t=\frac{2\Delta_1\Delta_2\sin^2\phi}{2\Delta_1\sin^2\phi+x^2\bar{\gamma}}$ and
 $u=\Big(\frac{2\Delta_1+x^2\bar{\gamma}}{2\Delta_1\Delta_2}\Big)t$, respectively,
a closed-form solution of $\mathcal{I}(x,\pi/2)$ is derived as
\begin{align}
\label{Iapiby2}
\mathcal{I}\left(x,\pi/2\right)
&=\frac{x\sqrt{2\Delta_1\bar{\gamma}}\exp\left(-\Delta_2\right)}{\pi (x^2\bar{\gamma}+2\Delta_1)}\Phi_1^{(2)}\left(1;1;1.5; \frac{2\Delta_1}{x^2\bar{\gamma}+2\Delta_1},\frac{2\Delta_1\Delta_2}{x^2\bar{\gamma}+2\Delta_1}\right),
\end{align}
where $\Phi_1^{(2)}(\cdot)$ is defined in (\ref{phi13}). 
\subsection{Closed-form Solution of $\mathcal{I}\left(x,\arctan(y/z)\right)$}
\label{AppendixAB}
Substituting, (\ref{mgf}) into (\ref{I}) followed by $t=\frac{2\Delta_1\Delta_2\sin^2\phi}{2\Delta_1\sin^2\phi+x^2\bar{\gamma}}$ and $u=\left(\frac{x^2(y^2+z^2)\bar{\gamma}+2\Delta_1y^2}{2\Delta_1\Delta_2y^2}\right)t$, respectively, a new closed-form solution of $\mathcal{I}\left(x,\arctan(y/z)\right)$ 
is derived as
\begin{align}
\label{akarctanba}
\mathcal{I}\left(x,\arctan(y/z)\right)
&=\frac{x\,y^2\sqrt{2\Delta_1\bar{\gamma}}\exp\left(-\Delta_2\right)}{2\pi (x^2(y^2+z^2)\bar{\gamma}+2\Delta_1\,y^2)}
\Phi_1^{(3)}\left(1;0.5,1;2; \frac{(x^2\,\bar{\gamma}+2\Delta_1)y^2}{x^2(y^2+z^2)\bar{\gamma}+2\Delta_1\,y^2},
\right.
\nonumber\\&\left.
\frac{2\Delta_{1}\,y^2}{x^2(y^2+z^2)\bar{\gamma}+2\Delta_1\,y^2},
\frac{2\Delta_1\Delta_{2}\,y^2}{x^2(y^2+z^2)\bar{\gamma}+2\Delta_1\,y^2}
\right),
\end{align}
where $\Phi_1^{(3)}(\cdot)$ is defined in (\ref{phi13}).

\subsection{Closed-form Solution of $\mathcal{I}\left(x,\arccot(y/x)\right)$}
\label{AppendixAC}
Substituting, (\ref{mgf}) into (\ref{I}) followed by $t=\frac{2\Delta_1\Delta_2\sin^2\phi}{2\Delta_1\sin^2\phi+x^2\bar{\gamma}}$ and $u=\Big(\frac{x^2(x^2+y^2)\bar{\gamma}+2\Delta_{1}\,x^2}{2\Delta_1\Delta_{2}\,x^2}\Big)t$, respectively, a new closed-form solution of  $\mathcal{I}\left(x,\arccot(y/x)\right)$ is derived as
\begin{align}
\label{Iapibyarctan}
\mathcal{I}\left(x,\arccot(y/x)\right)
&=\frac{x\sqrt{2\Delta_1\bar{\gamma}}\exp\left(-\Delta_2\right)}{2\pi ((x^2+y^2)\bar{\gamma}+2\Delta_1)}
\Phi_1^{(3)}\left(1;0.5,1;2; \frac{x^2\,\bar{\gamma}+2\Delta_1}{(x^2+y^2)\bar{\gamma}+2\Delta_1},
\right.
\nonumber\\&\left.
\frac{2\Delta_{1}}{(x^2+y^2)\bar{\gamma}+2\Delta_1},
\frac{2\Delta_1\Delta_{2}}{(x^2+y^2)\bar{\gamma}+2\Delta_1}
\right).
\end{align}
\section{Solution of $\mathcal{I}^H\left(x,\theta\right)$}
\label{AppendixB}
\subsection{Closed-form Solution of $\mathcal{I}^H\left(x,\pi/2\right)$}
\label{AppendixBA} 
Substituting, (\ref{mgfhighsnr}) into (\ref{Ihigh}), a solution of  $\mathcal{I}\left(x,\pi/2\right)$ is obtained as
\begin{align}
\label{IHapiby2}
\mathcal{I}^H\left(x,\pi/2\right)&=\frac{\sqrt{2\Delta_1}\exp\left(-\Delta_2\right)}{\pi x\sqrt{\bar{\gamma}}}.
\end{align}

\subsection{Closed-form Solution of $\mathcal{I}^H\left(x,\arctan(y/z)\right)$}
\label{AppendixBB}
Substituting, (\ref{mgfhighsnr}) into (\ref{Ihigh}), a solution of  $\mathcal{I}^H\left(x,\arctan(y/z)\right)$ is derived as
\begin{align}
\label{IHarctan}
\mathcal{I}\left(x,\arctan(y/z)\right)
&=\frac{\sqrt{2\Delta_1}\exp\left(-\Delta_2\right)}{\pi x\sqrt{\bar{\gamma}}}
\left(1-\frac{z}{\sqrt{x^2+y^2}}\right).
\end{align}

\subsection{Closed-form Solution of $\mathcal{I}^H\left(x,\arccot(y/x)\right)$}
\label{AppendixBC}
Substituting, (\ref{mgfhighsnr}) into (\ref{Ihigh}), a solution of  $\mathcal{I}^H\left(x,\arccot(y/x)\right)$ is got as
\begin{align}
\label{IHarccot}
\mathcal{I}^H\left(x,\arccot(y/x)\right)
&=\frac{\sqrt{2\Delta_1}\exp\left(-\Delta_2\right)}{\pi x\sqrt{\bar{\gamma}}}
\left(1-\frac{x}{\sqrt{x^2+y^2}}\right).
\end{align}

\bibliographystyle{IEEEtran}
\bibliography{reference}
\end{document}